# Seamless Flow Migration on Smartphones without Network Support

Ahmad Rahmati[1], Clay Shepard[1], Chad Tossell[2], Angela Nicoara[3], Lin Zhong[1], Phil Kortum[2], Jatinder Singh[3]
[1] Dept. of ECE and [2]Dept. of Psychology, Rice University, Houston, TX
[3] Deutsche Telekom R&D Laboratories USA, Los Altos, CA

## Abstract

This paper addresses the following question: *Is it possible to migrate TCP/IP flows between different networks on modern mobile devices, without infrastructure support or protocol changes?* To answer this question, we make three research contributions. (*i*) We report a comprehensive characterization of IP traffic on smartphones using traces collected from 27 iPhone 3GS users for three months. (*ii*) Driven by the findings from the characterization, we devise two novel system *mechanisms* for mobile devices to support seamless flow migration without network support, and extensively evaluate their effectiveness using our field collected traces of real-life usage. *Wait-n-Migrate* leverages the fact that most flows are short lived. It establishes new flows on newly available networks but allows pre-existing flows on the old network to terminate naturally, effectively decreasing, or even eliminating, connectivity gaps during network switches. *Resumption Agent* takes advantage of the functionality integrated into many modern protocols to securely resume flows without application intervention. When combined, Wait-n-Migrate and Resumption Agent provide an unprecedented opportunity to immediately deploy performance and efficiency-enhancing policies that leverage multiple networks to improve the performance, efficiency, and connectivity of mobile devices. (*iii*) Finally, we report an iPhone 3GS based implementation of these two system mechanisms and show that their overhead is negligible. Furthermore, we employ an example network switching *policy*, called *AutoSwitch*, to demonstrate their performance. AutoSwitch improves the Wi-Fi user experience by intelligently migrating TCP flows between Wi-Fi and cellular networks. Through traces and field measurements, we show that AutoSwitch reduces the number of user disruptions by an order of magnitude. In contrast, we show that brute-force switching would significantly increase user disruptions.

## 1. Introduction

Modern mobile devices have access to multiple networks. Not only do they have multiple network interfaces, such as cellular and Wi-Fi, but also a single interface may access multiple networks, such as Wi-Fi hotspots from different providers. Over time, for example as the user moves, the networks available to a mobile device and their qualities vary greatly. Many researchers have recently demonstrated the value of properly switching between networks [1, 2] or aggregating them [3, 4]. Switching between networks can significantly improve the performance [5, 6], energy efficiency [1, 7], and connectivity [8] of mobile Internet. In this work, we focus not on policies, but mechanisms for switching and/or aggregating networks on smartphones.

The key to switching between networks or aggregating them is to change the network for an existing flow without disrupting the corresponding application. Brute-force switching between networks, where one is simply disabled and another enabled, may lead to undesirable disruptions, as our own experience corroborates and also is confirmed by our user study. Solutions to this problem are available in the name of *handoff*. Some require infrastructure or home agent support, e.g. cellular handoff, connection gateway, and Mobile IP, which incur extra operating expenses and additional latency [9]. Others require changing the TCP/IP protocol, which has been shown to be practically very difficult. Not surprisingly, no automatic switching or aggregating solutions have been deployed in practice.

The important question this paper addresses is the following: On modern mobile devices, is it possible to seamlessly migrate TCP/IP flows between different networks without infrastructure support or protocol changes? Toward answering this question, this paper presents three research contributions.

First, we report a comprehensive characterization of network traffic on smartphones using three-month traces collected from 27 iPhone 3GS users. The characterization provides key insights into the motivation and rationale of our mechanisms. In particular, we have found that there are few concurrent network flows during interactive usage, flow lifetimes are typically short, and long-lived flows are often predictable.

Second, we present and extensively evaluate two novel system mechanisms implemented in a smartphone to migrate flows between networks without network support and without disruption to the user. The first mechanism, *Wait-n-Migrate*, takes advantage of the fact that TCP flows are short-lived. It establishes new flows on the new network, but waits for the pre-existing flows on the old network to terminate normally, up to a specific wait-time set by the migration policy. The second mechanism, *Resumption Agent*, leverages the resume function in modern servers and resumes a flow from wherever it was disconnected, in a manner transparent to applications. Based on our traces, we show that using Wait-n-Migrate, we can successfully migrate web flows for 90% and 95% of cases, for wait-time values of 10 and 100 seconds, respectively. With the addition of Resumption Agent, we show that for web flows that support resuming, we can virtually eliminate disruptions when switching between networks.

Third, we report an efficient implementation of the Wait-n-Migrate and Resumption Agent mechanisms on the iPhone platform, and show that their overhead is negligible. Based on the two system mechanisms, we further implement a sample network interface switching policy, AutoSwitch. AutoSwitch uses Wait-n-Migrate and Resumption Agent to offload data from cellular to Wi-Fi as much as possible, with minimum disruptions to the user. AutoSwitch using Wait-n-Migrate alone achieves over one order of magnitude reduction in disconnections in our real-life traces, e.g. from over 40% to well under 10% for 100 KB transfers while driving. Furthermore, when the content supports resuming, disruptions are almost entirely eliminated with the addition of Resumption Agent.

The rest of this paper is organized as follows. In Section 2, we present a motivational user study to show that brute force network switching is unacceptable to users, and then discuss related work. In Section 3, we present the characterization of network traffic on 27 iPhone 3GS users and provide insight to the characteristics of network flows on modern smartphones. Based on these findings, in Section 4, we present the design and trace-based evaluation of Wait-n-Migrate and Resumption Agent. In Section 5, we report their implementation on iPhone and evaluate their performance impact. In Section 6, we present an example application, AutoSwitch, of the resulting seamless flow migration without network support. Finally, we discuss methods to further enhance our mechanisms for increased performance in Section 7, and conclude in Section 8.

## 2. Background
### 2.1 Consequences of Brute-Force Switching

Without network support, smartphones switch between networks (e.g. cellular and Wi-Fi) in a brute-force manner: they terminate all application flows on the old network and enables the new network. This behaviour is shared across all the three major smartphone platforms we studied; iOS, Android, and Windows Mobile[1]. It is then up to the application, or often the user, to detect the disconnection and retry over the new network This brute-force switch introduces disruptions to interactive sessions. According to our personal experience, network disruption is noticeably annoying, and particularly prevalent for large web pages or during poor connectivity. To better understand the usability impact of network disruption (e.g. as will be experienced due to brute-force switching), we performed a formal user study with 10 participants from the Rice student community who already used Internet-ready smartphones. The study included an equal number of males and females and four participants with non-engineering backgrounds.

Our study consisted of two parts. The first part asked the users to open a copy of a regular news website cached on our server for consistency. We then asked users to perform a number of text identification tasks on three individual pages. The participants were later directed to a cached copy of a mobile news search engine, where they were asked to identify several stories and their sources. During the study, our server automatically disrupted the data flow for the first load of three of the five page loads. The users had to refresh their browser to completely load each page. This simulated the impact of a brute-force migration. Participants were free to either use their own phones or our iPhone for the purpose of this study.

For the second part, we interviewed the participants to assess their browsing experience, including several questions on a 1 – 5 Likert scale (agree – disagree), and several open ended questions. All 10 participants agreed or somewhat agreed that disruptions are an *annoying experience*. Interestingly, all 10 also agreed or somewhat agreed that they *have had similar experiences prior*, and that they *typically refresh a page that has failed to completely load*.

During the open ended question sessions, when asked whether they have experienced this phenomenon more often in specific web sites, 9 of 10 mentioned that they experience it more frequently with larger transfers, e.g. mentioning pages that are as "heavier" or "with lots of graphics". When asked whether they have experienced this phenomenon more often in specific conditions, 8 of 10 correctly identified that they experience it more frequently during one or more network conditions (e.g. low signal, moving). We can see that even without intentional network switching, users are subject to unwanted and annoying network disconnections.

While our user study was conducted with a small number of participants (n=10), considering the high confidence intervals, our findings are expected to be true with the majority of user populations similar to our participants. For example, the 90% Agresti-Coull confidence interval [10] for 8 and 10 positive answers out of 10 are (0.52 , 0.91) and (0.66 , 1), respectively, i.e. there is a 90% chance that the statistics for the population falls in those intervals.

In summary, we confirmed that network disruptions annoy users. We also found that typical users have extensive experience with network disruptions, and have even figured out the conditions in which they often occur. A successful solution to for network disruptions must not blatantly change the user experience or discard the partially received content. These findings motivate and assist both the design of our mechanisms and our example application, AutoSwitch.

---
[1] The only exception was iOS and only when switching from cellular to Wi-Fi, where it keeps existing connections indefinitely on their original interface.

## 2.2 Related Work

TCP/IP lacks built-in support for switching between multiple networks (handoff) or aggregating their throughput (multihoming). Therefore, there exists a body of research on providing session continuity [11] between different networks, i.e. maintaining the same IP address while moving between networks. Current solutions for session continuity fall into three categories. First is to have one network as the slave to a master network, where all traffic is directed through the latter [12], as in Virtual AP. However this requires unified management of the networks, increases traffic on the master network, and increases latency. The second category of solutions utilize a mobility gateway in the infrastructure [13, 14], to act as a proxy between a mobile device and the Internet. For example, such gateways have been employed for switching between interfaces (Wiffler [5]), for multihoming ([3, 4, 15]), and for striping ([13, 16-23]). However, routing all flows through a fixed gateway can increase the connection latency. The third category of solutions modify or extend the TCP/IP protocol support for mobility, e.g. by adding explicit support, as in [24, 25], or through Mobile IP [26-28], where a home router or agent handles mobility and packet forwarding. However, the extra forwarding increases the traffic on the home agent and more importantly, the extra distance travelled by packets increases the connection latency. Mobile IPv6 eliminates the need for a specific foreign agent, but in return requires individual mobile nodes to perform the forwarding operations, with similar drawbacks.

All three categories of solutions discussed above require additional infrastructure or network support, and thus are not immediately deployable. Those that have begun deployment suffer from limited or unsuccessful adoption. Furthermore, these solutions increase network latency, which is already known to be a major bottleneck in mobile Internet performance [29]. In sharp contrast, we present and evaluate two novel and complementary switching mechanisms that can be fully implemented on mobile devices without requiring network or application support, and with insignificant additional latency. This allows our techniques to be deployed immediately without changes to applications or infrastructure.

There are two solutions related to *Resumption Agent*. Resuming static content is typically supported by download managers such as *wget*. Yet, most other applications, e.g. browsers, lack resume functionality. Snoeren et al. [29] supported resumption through a client agent for the purpose of failover between replica servers, while keeping servers largely unchanged. In contrast, Resumption Agent is an application agnostic solution for network switching and provides automatic resuming capabilities for all pre-existing applications. Furthermore, it can handle the challenges of dynamic content and secure HTTPS connections.

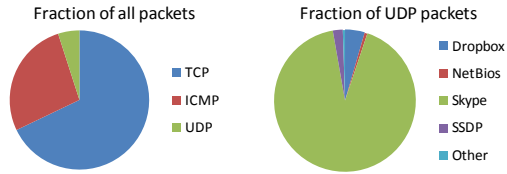

**Figure 1: Fraction of packets for each protocol (left), and fraction of applications for UDP packets (right).**

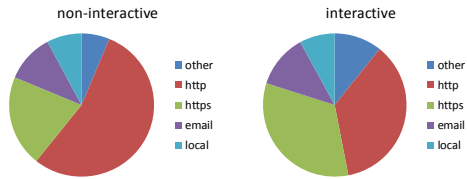

**Figure 2: Fraction of TCP flows for each application type (Left: non-interactive sessions. Right: interactive sessions, i.e. phone display was on).**

Recently Alperovich and Noble have proposed to improve Wi-Fi performance for PC clients by switching and balancing connections between multiple Wi-Fi access points (APs), e.g., as enabled through Virtual Wi-Fi [30]. They also retain pre-existing connections on their original AP, while assigning new connections to new APs. Yet, our work focuses on smartphones and presents mechanisms for switching between multiple, heterogeneous networks. We go well beyond retaining pre-existing connections by addressing long-lived flows and supporting pre-existing applications on mobile phones.

Finally, there have been several studies addressing smartphone usage and network traffic characteristics [32, 33]. Our contribution in traffic characterization compliments these works, in particular for the purpose of migrating flows between networks, by providing detailed analysis of traffic protocols, flow length and concurrency, and the active application concurrent to the flows. These findings are crucial for designing and evaluating the feasibility of network migration.

## 3. Network Flow Characterization

A thorough understanding of the characteristics of network flows on modern mobile devices is critical to the seamless migration of flows. We next report a first-of-its-kind study based on detailed network flow traces from 27 iPhone 3GS users. The characterization provides key insights for our design, as described in Section 4.

### 3.1 iPhone Field Trace Collection

We gathered real-life network traces from 27 iPhone 3GS users over the course of 3 months by installing logging software we developed, called LiveLab [34]. We chose the iPhone because it represents the cutting edge of smartphone design for usability, accounting for 55% of all mobile internet traffic in the US as of October 2009 [35]. Additionally, iPhone users have access to the largest number of third-party applications from the Apple App Store and numerous third-party repositories.

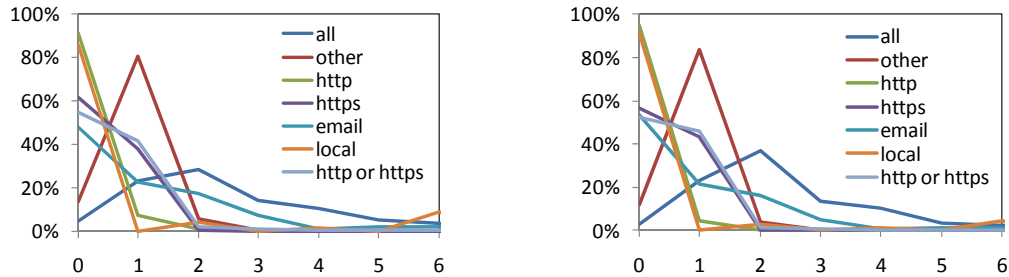

**Figure 3: Distribution of the number of concurrent TCP flows for different TCP ports, average among all users (Left: non-interactive sessions. Right: interactive sessions.)**

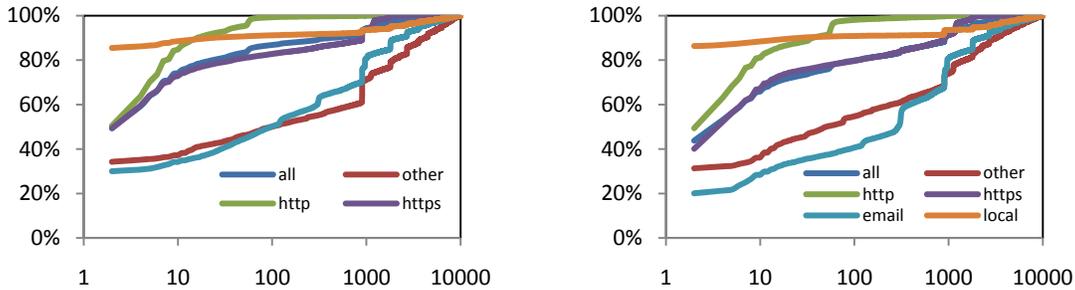

**Figure 4: CDF of TCP flow lifetimes in seconds, based on TCP port, average among all users (Left: non-interactive sessions. Right: interactive sessions.)**

Whenever the phone's CPU is not asleep, LiveLab records TCP network connection statistics every two seconds using the *netstat* tool, also available on Windows and Linux/Unix platforms. Moreover, LiveLab records the application being used and the display status in real time, and Wi-Fi signal strength for the currently connected AP and all visible APs every two seconds and 15 minutes, respectively. Finally, it recorded the complete packet headers for three of the participants over one month, in order to gauge the data flow over UDP. We refrained from deploying this packet-level logging for longer time or more users due to its overhead. The data is recorded on the phones, and is transferred nightly to our servers in a secure fashion.

While our participants were not recruited to accurately represent the vast mobile user population, the data collected from them provides an unprecedentedly detailed look into the connectivity on contemporary mobile devices.

### 3.2 Focus on TCP Flows

The packet-level logging data shows that out of the three common IP protocols in use TCP, UDP, and ICMP, TCP flows present the main challenge towards flow migration. TCP, ICMP, and UDP account for 68%, 27%, and 5% of all packets, respectively (Figure 1). While we will examine TCP flows in details later, we will first discuss ICMP traffic and UDP flows.

ICMP packets are typically not used by interactive applications, but by devices to for diagnostics, device discovery and error messages specific to each network. Therefore, for the purpose of switching between networks, they can be safely ignored.

UDP flows only contribute 5% of the total packets. Yet, we analyze the UDP flows based on port numbers, and further corroborate this analysis with the applications being used. Notably, the phones were almost always listening to all UDP ports. We have found the following services and applications utilize UDP on the phones (Figure 1):

- Skype (92%) uses UDP ports 12340 and 20515.
- Dropbox (4%) uses UDP broadcast on port 17500
- Simple Service Discovery Protocol (SSDP) (2%) is used to advertise and discover network services.
- NetBIOS (1%) for local area network device discovery and networking
- Other (<1%) such as NAT Port Mapping

With the exception of Skype, all of these are network and discovery services and specific to a particular network. Therefore we will ignore them for the purpose of switching between networks, similar to ICMP traffic. We will analyze how Skype can be migrated to a different network in Section 3.4.1. For the remainder of this paper, we will focus exclusively on TCP flows unless mentioned otherwise.

#### 3.2.1 TCP Flows

Using the port number of the server, we divide external TCP flows into three categories:

- Web (HTTP: 80, HTTPS: 443): These are used by not only the browser, but also by a number of native applications that utilize web services or a built-in browser.
- Email (IMAP: 143, 993, POP3: 110, 995, SMTP: 25, 465): These are used by the native email client, and will not include email accessed through the browser.
- Other: All other applications and services.

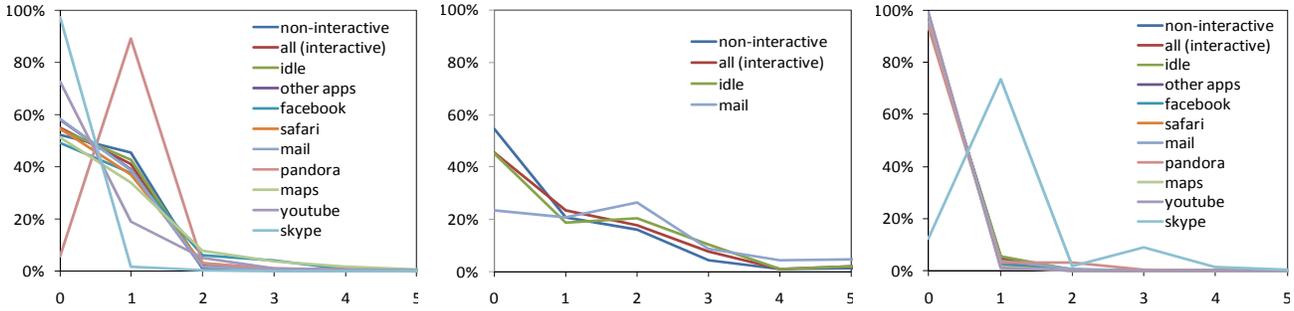

**Figure 5:** Distribution for the number of TCP flows when running different Internet applications. (Left): Web. (Center): Email. (Right): other ports, excluding the Push Service.

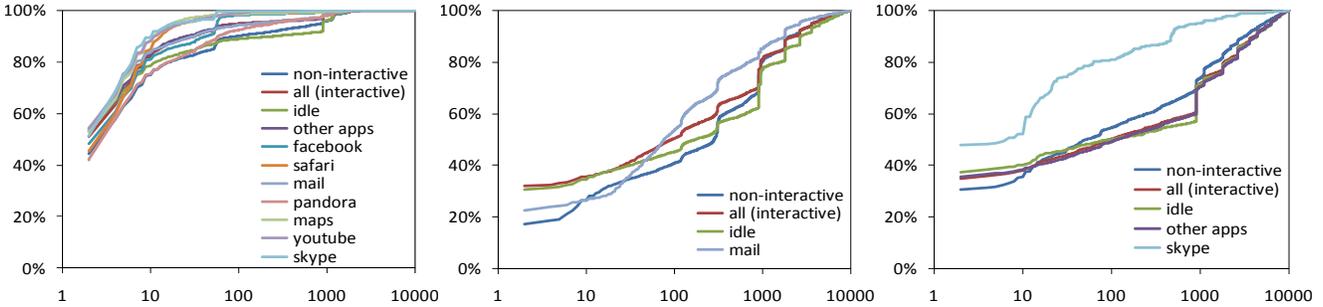

**Figure 6:** CDF of TCP flow lifetimes (seconds), based on active application. (Left): web ports. (Center): email ports, (Right): other ports.

Figure 2 shows the fraction of TCP flows utilized for each application during both interactive and non-interactive usage. We use the display status (on) as an indicator of the phone is being used interactively. We can see that more than three quarters of TCP streams are web flows, highlighting the importance of handling them properly. We also separate and ignore local (loopback) flows that reside only on the phone.

Figure 3 and Figure 4 show the distribution of the number of TCP flows and the CDF of flow lifetimes, respectively, for both interactive and non-interactive sessions among all users. We can see that flows have similar characteristics during interactive and non-interactive usage, yet, on average, flows during interactive usage have slightly shorter lifetimes. In the following Sections, we will study them in further detail, according to application use.

### 3.3 Flow Concurrency

While analyzing the LiveLab data, we were surprised to discover that there are few concurrent flows on the iPhone platform. However, there almost always exists one particular flow, 97% of the time that the phone is awake. We have identified that flow as *Apple's push notification service, on port 5223*. The median number of flows was 2 for both interactive and non-interactive sessions. Figure 5 shows the distribution of number of concurrent TCP flows, excluding the Apple Push service, whenever the phone's CPU was running for the three port types presented in Section 3.2.1 (web, email, other). We identified the top seven applications that require Internet access using the data from our field study, which include Pandora (music streaming) and Skype (instant messaging, voice over IP). These applications account for over 95% of interactive phone Internet use. Non-interactive usage, including when the display was off, idle time, when the home screen was displayed are presented separately. Other applications, including those without specifically requiring internet connectivity, are clustered together as others. For email and other ports, we display only the applications that we have determined to use those ports.

We can see that even when running internet enabled applications, the phone is rarely engaged in multiple TCP flows simultaneously. The small numbers of simultaneous TCP flows shows that *for web applications on mobile phones, multihoming mechanisms (i.e. non-striping) are effective for at most 20% of flows*, as the other 80% of times when a web flow exists, it is a single flow. However, we expect this number to increase as more applications and services on mobile devices become available. The mail application, while not typically data intensive, presents an exception, as it regularly uses multiple flows when active.

### 3.4 Flow Lifetime

We have found that most interactive flows on the phone were short lived, and it is often possible to automatically predict long-lived flows. We measure the flow lifetime without including the connection / teardown phase (e.g. wait_fin). Our logs show a wide variation in the lifetime of TCP flows on the experimental phones, in particular between interactive and non-interactive usage sessions.

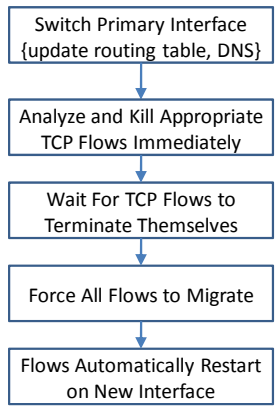

Figure 7: Flowchart for Wait-n-Migrate

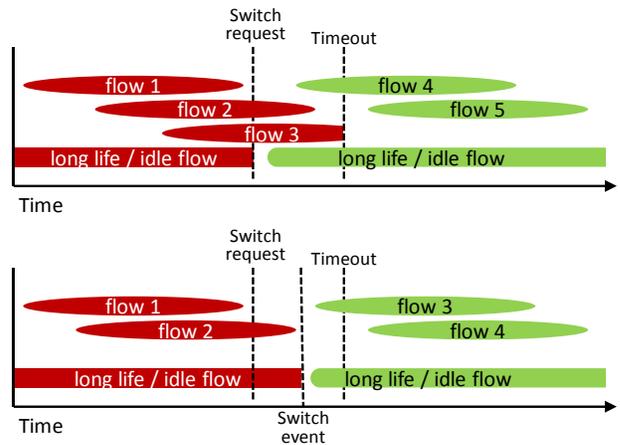

Figure 8: Wait-n-Migrate operation (Top), and the special case without requiring simultaneous connectivity (bottom).

Figure 6 shows, on average among our participants, the CDF (cumulative distribution function) of TCP session lengths for different TCP ports and different active applications, similar to Section 3.3

Our first finding is that *most flows are short lived*. In fact, 50% and 44% of flows for non-interactive and interactive sessions, respectively, are ~2 seconds or less. In turn, *this limits the effectiveness of power saving schemes which rely on long-lived downloads*, such as CatNap [36].

Our second finding is that it is possible to predict flow lengths based on active application and port, i.e. the distribution of flow lifetimes varies significantly based on TCP port, active application, and whether the phone is being used interactively. For example, as shown in Figure 6, the fraction of short lived email flows (i.e., IMAP, SMTP, POP3) is much lower: 30% and 20% for non-interactive and interactive sessions respectively. Similarly, the Apple Push service is known to be long lived. On the other hand, as shown in Figure 6, TCP flows during web browser sessions were shorter than average. We will later see how these findings are important for our switching mechanisms.

### 3.4.1 Long-lived Non-Standard TCP Flows

We next consider long-lived flows that use non-standard protocols, other than web, ftp, and email. Such flows are difficult, if not impossible, to migrate without network support. However, a close examination reveals that such flows usually do not require migration support at all.

First, long-lived TCP flows based on closed application protocols are usually from background, non-interactive applications. Therefore, their disruption or brute-force migration will be unnoticeable to users.

More importantly, the handful of applications that utilize long lived non-standard protocols already provide support to migration in various forms because the application developers anticipate the possibility of disconnection. For example, applications such as Push notifications, Twidroid, and many instant messaging applications are designed to gracefully and automatically re-establish a connection after being disconnected. Another example, Pandora, a common Internet radio streaming application, and the only one that appeared in our participants' list of top 25 applications, skips the unbuffered part of the current song, i.e. at most suffer skipping part of a track. For yet another example, as long as the primary interface in the system routing table is correctly updated, e.g. as is the case with our mechanisms or when the user manually enables or disables Wi-Fi, Skype switches to the new network for both its TCP and UDP connections, without dropping a call and with only a very short period (~1 sec or less) of muting in the audio. However, if the system is unaware of the disruption (e.g., moving out of Wi-Fi coverage), Skype will drop the call. This highlights the importance of notifying applications and the system of the network change.

### 3.5 Background Applications

While the iPhone 3GS we used in the study was the state-of-the-art phone at its time, it lacks official support of multitasking for third-party applications as of OS 3.x. Yet, we expect that increased multitasking will not reduce the usability and effectiveness of the Wait-n-Migrate and Resumption Agent mechanisms. We note that Android and the newly released iPhone iOS 4.0 allow background applications, e.g. Skype and Pandora, to access data networks [37]. This, alongside the increasing processing power and memory of phones, suggests an increase in the usage of background capable applications (e.g. instant messaging, Twidroid). Therefore, we would expect to see an increase in the number of simultaneous network flows, from those shown in Figure 3 and Figure 5.

Assuming the device can remain connected to two networks simultaneously, we can consider each application independently for both Wait-n-Migrate and Resumption Agent. Indeed, flows belonging to a certain application do not impact other application flows. Therefore, an increase in the number of multitasked applications will not affect the general performance of our mechanisms.

## 4. Migration without Network Support

Based on the findings from Section 3, for the purpose of migrating network flows between networks, we focus on seamlessly migrating short lived flows or flows using standard protocols such as HTTP and FTP. We provide two novel and complementary mechanisms for migrating such flows without network support. We envision that in most systems, Wait-n-Migrate will be used primarily, and Resumption Agent will be used to migrate flows that were not successfully migrated by Wait-n-Migrate.

### 4.1 Wait-n-Migrate

Our first method leverages the fact that most flows are short lived, as seen in Section 3. Wait-n-Migrate typically requires the device to be able to connect to multiple networks simultaneously. This may be through multiple interfaces (e.g., 3G and Wi-Fi) or through one interface (e.g., multiple Wi-Fi networks through Virtual Wi-Fi [31]).

In order to migrate one or more flows between two networks, Wait-n-Migrate operates as follows (Figure 7). First it enables both networks so the system has simultaneous connectivity to both. Second it ensures all new flows are created on the new network. Then it waits for the flows on old network to terminate naturally, up to a specific wait-time (Figure 8). The wait-time for each flow is a parameter determined by the particular migration policy and can be set according to application, bandwidth, and power considerations, and may be adaptive according to flow characteristics presented in Section 3. Different wait-time values can be used for switching to different networks. For example, when the system policy requests a network switch to a slower or less efficient network, e.g. in order to assure connectivity, the wait-time can be set to infinite, i.e. until losing connectivity. On the other hand, when switching back to the faster / more efficient network, a shorter wait-time should be used. Finally, if there are no remaining flows on the old network, the system can disable or power it off altogether.

When the system cannot be connected to both networks simultaneously, a special case of Wait-n-Migrate can be used. This special case takes advantage of the fact that that most TCP flows are short lived. It monitors TCP flows and attempts to choose the best moment to switch within a specifically allowed time range, in order to minimize disruptions. This is possible through the statistical properties of TCP flows, as presented in Section 3.

Finally, Wait-n-Migrate can employ flow lifetime prediction to further improve its effectiveness and efficiency. Wait-n-Migrate does not interfere with short-lived flows in order to avoid user disruption. However, for flows that are known to be highly likely to live beyond the wait time, e.g. based on the findings in Section 3, Wait-n-Migrate can terminate them immediately. For example, we already know that several types of flows are long lived, e.g., Push notifications and idle email flows. If the device is switching to a faster or more energy-efficient network, Wait-n-Migrate can terminate such flows immediately, thus improving performance.

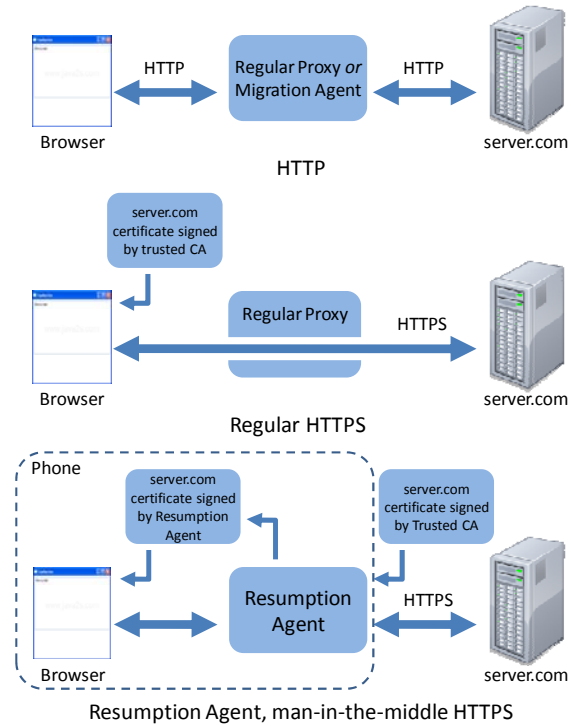

**Figure 9: Regular proxy operation and Resumption Agent man-in-the-middle operation for a browser application**

### 4.2 Resumption Agent

Our second method, Resumption Agent, leverages the fact that many interactive applications use standard application layer protocols such as HTTP, HTTPS, as highlighted in Section 3.2, and that most servers for these protocols support resume. Resumption Agent is a locally run proxy that enables flow migration for most such flows. It provides a safety net to reduce the user impact of network switching when Wait-n-Migrate terminates a flow for migration. With Resumption Agent, Wait-n-Migrate can be more aggressive in migrating flows and therefore allow for faster switching.

Resumption Agent can support any application that allows resuming from a specified location in the transfer. Several key standard application-layer protocols, including HTTP and FTP, provide adequate support for resumption of a terminated transfer. For example, the HTTP standard, from version 1.1 onwards (1996), supports specifying a *range* when requesting a web page. The FTP standard also supports resuming via the *rest* command. Standard email protocols (e.g. IMAP, POP, and SMTP) can also be restarted from the beginning of any email, or any individual attachment in the case of IMAP.

Resumption Agent works as follows. It requires a background service running only on the device itself, which acts as a proxy, and modifies the phone settings so that applications use this proxy to connect to the internet. If a flow is disconnected prematurely, Resumption Agent automatically resumes the transfer from where the flow was cut off. Therefore, when a flow needs to migrate to a new network, it can be terminated on the old network and resumed on the new network in transparent manner to the application. Finally, Resumption Agent can employ flow lifetime prediction to further improve its effectiveness and efficiency. For web flows, their sizes are typically know at the beginning of the transfer, through the HTTP header response *Content-Length*. If Resumption Agent is used in conjunction with Wait-n-Migrate, the content length and bandwidth can further assist in determining whether to kill flows immediately or wait for them to terminate normally.

We note that download managers, such as *wget*, support the automatic resuming of static content. Yet, they are unable to handle the challenge of unsupported content, as discussed in 4.2.1. More importantly, web browsers (on both PCs and phones), and most other applications (e.g. the iPhone YouTube application) lack automatic resuming functionality. In contrast, Resumption Agent is application agnostic and appears as a regular proxy server to applications, thus providing a system level solution for all pre-existing applications. Moreover, Resumption Agent can handle network migration and two non-trivial challenges to Resumption Agent for web flows, posed by unsupported content and encrypted HTTPS flows. We next discuss them and present our solutions.

### 4.2.1 Unsupported Content

There are three groups of content that cannot be resumed in the middle of the transfer.

(*i*) The first group includes content that does not allow resuming. For example, some servers may ignore HTTP Range requests altogether or for specific content, such as small transfers, or chunk encoded data (the size of the data is not known beforehand). In this case, the transfer, if interrupted, must be restarted from the beginning, resulting in a second and unnecessary transfer of the initial portion, which the Resumption Agent will ignore.

(*ii*) The second group is content uploads, usually using HTTP POST, in which there is always the risk of repeating an action, e.g. a purchase. In such cases, such as when the user refreshes a page with POST content, web browsers present the user with a warning. Resumption Agent uses the same behaviour and will avoid automatically resuming such a transfer if it is disconnected.

(*iii*) The third group is dynamic content that changes significantly for every reload. Resumption Agent deals with dynamic content using two methods. First, the HTTP headers *Pragma:no-cache* and *Cache-Control:no-cache* in the request and response headers, respectively, indicate dynamic content, as the prevent proxies and other web servers from caching the content. Thus, if Resumption Agent sees these tags, it can abstain from automatically resuming a failed transfer. Second, in order to support dynamic content that does not provide hints in the headers, Resumption Agent always resumes from a preset length prior to the disruption. It then compares the overlapping sections. If the overlapping sections are identical, Resumption Agent will simply continue with the resume. If the overlapping sections become identical after applying a small offset to the data, e.g. to account for a slightly smaller or larger dynamic advertisement content, it will correct the offset and can continue with the resume. Only if the overlapping sections are not identical even after applying an offset, will Resumption Agent abort the resume and the transfer will fail.

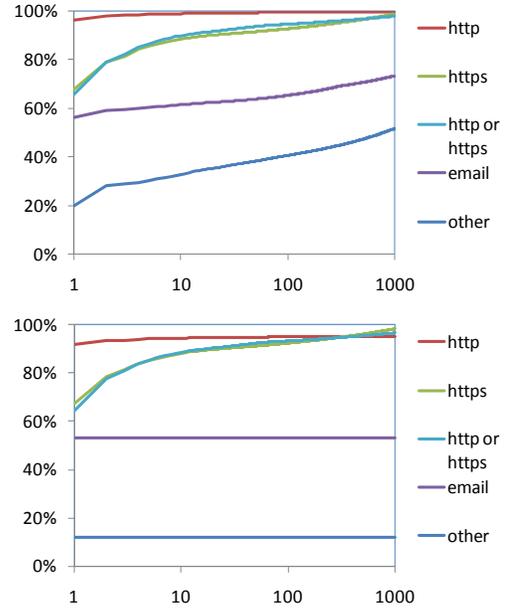

**Figure 10: Performance of Wait-n-Migrate (top) and the special case without simultaneous network connectivity (bottom), measured as the percentage of flows successfully migrated to the target network, for different timeout values using our field-collected traces.**

### 4.2.2 Encrypted HTTPS Flows

A greater challenge comes from HTTPS, as it is impossible for a proxy to directly inspect its contents, which is end-to-end encrypted by SSL. Indeed, when an application wants to connect to a HTTPS server through a typical proxy, it sends a CONNECT command to the proxy. The proxy, upon validating the eligibility request, will create a tunnel to the requested server, without touching the transferred content. Such a configuration, with end-to-end encryption would make it impossible to analyze the data, necessary for transparently resuming or striping transfers.

Resumption Agent employs a novel and elegant two-part solution to this challenge. First, it will exploit a *man-in-the-middle attack*. That is, as shown in Figure 9, Resumption Agent presents itself to the client as the destination server. It then connects to the destination server, and therefore has access to the transferred stream, and can perform the same functionality it does for HTTP. We note that the open source web proxy, *squid*, has built-in support for such man-in-the-middle operation [38].

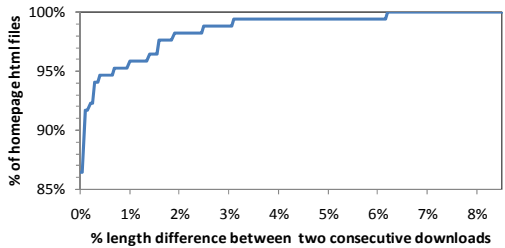

**Figure 11: Non-static web pages often have the same or similar content lengths: CDF of the length differences of two consecutive downloads (among the top 100 pages accessed by our users)**

A standard man-in-the-middle attack by a third party is, however, unable to present the correctly signed certificate to the client application, and depending on system policies, it typically raises a warning to the user. Changing system policies to ignore security certificates would open the door to any man-in-the-middle attack, and is therefore unacceptable. Indeed, in order to maintain security, the certificate check must be strictly enforced.

The second part of our solution addresses this challenge. All computer systems, including our iPhones, depend on a number of preinstalled Certificate Authorities (CAs) to sign and validate all server certificates. Since Resumption Agent is *not* a third party, it can install its own local CA on the device, without compromising security. This is possible on the iPhone [39] as well as other platforms. Resumption Agent can then sign the certificates it presents to applications, preventing them from displaying warning messages. Resumption Agent has to create a new certificate once for each HTTPS domain the user accesses. We have measured the overhead of certificate generation on the iPhone 3GS to be on average 1.7 seconds, with a standard deviation of 1.2 seconds, measured over 100 experiments. Furthermore, to completely avoid this latency, the device can use the typical CONNECT command the first time the user accesses a new site, but generate the certificate for subsequent accesses.

When connecting to a server, Resumption Agent verifies the security certificate of the server instead of the application (e.g. browser). In order to maintain security, Resumption Agent (instead of the application) displays a warning to the user if a server's certificate is not correctly signed. The user can then decide whether to continue or forgo a potentially unsecure connection. In order to maintain security it is imperative to strictly enforce the certificate verification between clients and servers. We conjecture that a consistent warning for invalid certificates may be more understandable to end users than application specific warnings. Therefore, Resumption Agent can in fact reduce bad decisions by users and increase security.

### 4.3 Trace-Based Evaluation

In this section, we demonstrate the efficacy of Wait-n-Migrate and Resumption Agent using our field collected traces of real-life usage.

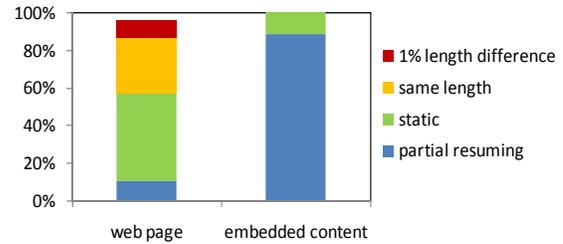

**Figure 12: Most web pages and all of their embedded content (among our top 100 pages) are supported by Resumption Agent**

To evaluate Wait-n-Migrate, we calculate the percentage of flows that are successfully transferred between networks for different wait-time values, assuming a time uniform probability of the system attempting a switch. To evaluate Resumption Agent, we measure the feasibility of resuming video streaming and browsing, and show that both YouTube and the majority of the websites participants most commonly visited indeed support resuming.

*4.3.1 Wait-n-Migrate*

As mentioned in Section 4.1, Wait-n-Migrate requires both networks/interfaces to be connected simultaneously, at least for the duration of the migration. For this evaluation, we assume the device intends to migrate all existing flows to a new network. We have used our traces to calculate the percentage of flows that Wait-n-Migrate can successfully migrate to the new network without disruption, shown in Figure 10. For example, Wait-n-Migrate successfully migrates all web flows for 90% and 95% of cases for wait-time values of 10 and 100 seconds, respectively.

As discussed in Section 4.1, there is a special case of Wait-n-Migrate that is employed when the system can only remain connected to one network, which waits for the moment where there are no ongoing flows to switch the network. For our evaluation, we assume that this special case waits for the moment when there are no web flows to switch between networks. We have used our traces to calculate the percentage of flows that the special case of Wait-n-Migrate can migrate to the new network without disruption in this manner, shown in Figure 10. Since our policy does not wait for presumably non-interactive flows (i.e. non-web) to end, we can see a significantly larger number of disconnections for those flows. Yet, the special case of Wait-n-Migrate performs relatively close to Wait-n-Migrate for web flows, as there are rarely multiple web flows in our traces, as shown in Section 3. However, we believe that increased complexity and multitasking in future applications will increase the performance difference of Wait-n-Migrate and its special case.

*4.3.2 Resumption Agent*

We have studied the applicability of Resumption Agent for two important applications, the web browser and the YouTube application.

We have tested YouTube and it is fully supported by Resumption Agent; the stream is based on standard HTTP

protocols and our experiments show that YouTube servers indeed support resuming videos at an arbitrary location.

We evaluate the applicability of Resumption Agent for web browsing by identifying whether it can be effective for the top 100 websites our users have visited. We used our user study logs to generate the list of top 100 websites our users visit. For each of these 100 sites we measure the resume capability of the website's homepage and its embedded media (e.g. images). We test the homepages (i.e. top page) since we found that many deeper, pages may depend on previous state information, e.g. a specific referrer, cookies, or user login). We crawl these sites both as an iPhone browser and as a desktop browser, set through the *User-Agent* HTTP header. Every crawl, we download each item three times, twice in full, and once from the middle of the transfer to determine 1) if the item supports resuming, and 2) if the item is static. We present the results for the iPhone and desktop browsers together, since they were similar.

As shown in Figure 12, we found that 100% of embedded media is static and therefore supported by Resumption Agent. 91% of those support resuming from the middle of a transfer; i.e. without the need to re-transfer the already transferred part. Among the HTML homepages, 57% were static, and 9% had the same content length between our two consecutive downloads, but had slightly different content. Furthermore, most others had content lengths very close to each other. Figure 11 shows the Cumulative Distribution Function (CDF) for the length differences between two consecutive downloads for our top 100 pages. We can see that another 30% had content lengths within 1% of each other. Therefore, we expect them to be supported by Resumption Agent, as described in Section 4.2.1. We note that only 16% of the HTML pages can be resumed from the middle of the transfer, vs. 89% of the embedded content. The remaining pages that do not support HTTP resume functionality incur an extra overhead of re-downloading the already transferred part, but can still be resumed transparently to the application.

Finally, while none of the embedded content used HTML tags to disallow caching, we observed that 30% of the HTML pages were marked as such. However, of the HTML pages that disallowed caching, 36% in fact had static content, and 44% had content with the same length. Therefore, we conjecture that the no-cache response header may possibly be ignored by Resumption Agent.

# 5. iPhone 3GS based Implementation

We implemented both the Wait-n-Migrate and the Resumption Agent mechanisms on the iPhone platform and measured their system overhead to be negligible. While the iPhone is a closed platform, a jailbreak has been consistently available, making it possible to develop low-level system software and implement these mechanisms.

We have constrained our solution to support legacy applications. The methods we have used, e.g. to acti-

```
Establish Connection to Server
Forward Client Headers
Forward Server's Reply Headers to Client
While download_unfinished && tries < MAXRETRIES  {
    Forward CHUNKSIZE bytes of data
    If disconnect_signaled
        Disconnect connection to server
    If disconnected && eligible for resume
        Reconnect to server
        Forward original headers
            If server supports ranges
                Request range starting at prior to disconnect
            else
                Download and discard previously downloaded data
    Tries++
}
Close sockets
```

**Figure 13: Pseudocode for Resumption Agent**

vate/deactivate network interfaces, are supported on every major OS without kernel modification, however some implementation details are OS specific, as described below.

## 5.1 Wait-n-Migrate

The implementation of Wait-n-Migrate realizes four functions: monitoring flows, selecting the primary network interface, terminating individual TCP flows, and disabling a network interface:

(*i*) *Flow Monitoring*: An intelligent network switching policy requires detailed knowledge of flow properties. For example, it may want to force the migration of high bandwidth flows with long durations immediately while switching to Wi-Fi. Towards this end, Wait-n-Migrate continuously records flow statistics, such as application, duration, destination, and bandwidth, for all flows. This information is reported in real-time, as well as kept in a database which is made available to the switching policy.

(*ii*) *Selecting Primary Network*: The implementation of Wait-n-Migrate depends on the ability to modify the system's routing tables to direct all new flows through the new network. The routing table consists of a set of prioritized rules dictating which interface and gateway to use for establishing outgoing sockets. All common operating systems have a routing table which they allow to be modified through well documented system calls. While it is possible to directly modify the routing table on the iPhone, we found that any modification to the primary default gateway triggered the system to reset the routing table. Instead we were able to use the *scutil* command to change the priority of the networks; this in turn automatically changes the routing table appropriately, as well as the DNS settings, and sends a system wide notification of the network change (as it typically does when switching interfaces). The overhead for invoking a switch is quite small, as it simply changes a system setting, and takes less than 300ms to complete. scutil is an OSX specific tool, though other OSs provide proprietary methods to select the primary network interface. Conveniently, the iPhone does not disable the cellular

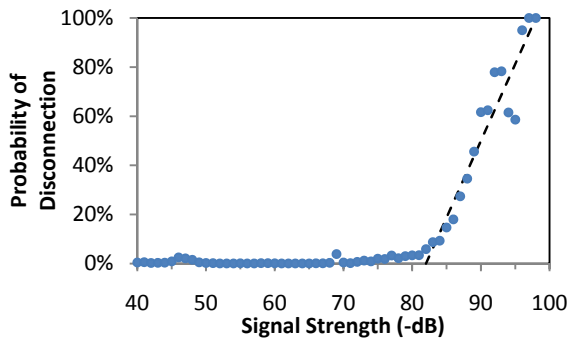

Figure 14: Probability of disconnection vs. Wi-Fi signal strengths

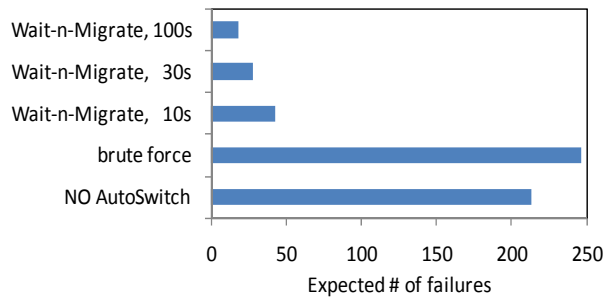

Figure 15: AutoSwitch significantly reduces the expected number of disruptions in 1120 hours of interactive usage traces with Wi-Fi enabled, for different wait-time values

interface while Wi-Fi is connected, as some platforms, such as Android, do. This, however, does not have a power impact as the phone must leave the cellular interface on in order to receive calls.

(*iii*) *Terminating Flows:* As previously mentioned, it may be necessary to force the migration of specific flows, such as those that are known to be of long duration. We have achieved this by porting *tcpkill* to the iPhone platform. tcpkill has been ported to all major kernels, including Darwin, Windows, FreeBSD, OpenBSD, HP-UX, AIX, Solaris, and Linux. tcpkill uses *libpcap* and *libnet* to detect/monitor the TCP stream and inject a TCP RST packet which kills the connection. When the application reconnects it is automatically routed through the new network.

(*iv*) *Disabling Network*: Wait-n-Migrate provides a mechanism to disable the entire network being migrated from. This can be useful after individual flows have been appropriately dealt with, or none of the flows require special treatment, depending on policy. Every major OS has methods to disable network interfaces. For UNIX based OSs this is typically "*ifconfig* interface down". Unfortunately this method currently does not work on the iPhone, however similar functionality can be achieved through the *scutil* or *ipconfig* commands. Additionally, low level *ioctl* calls can also accomplish this behaviour.

## 5.2 Resumption Agent

We implemented Resumption Agent in 1400 lines of C code; it can be built and run on any POSIX compliant system, including Linux and iOS. Resumption agent is similar to other proxies, such as squid, in that it acts as a relay point for Internet communication between clients and servers, complies with HTTP 1.0/1.1 specifications, and handles multiple concurrent connections.

Our implementation (Figure 13) leverages standard UNIX sockets and multithreading. When Resumption Agent starts it initializes a pool of worker threads, using *libpthread*, to handle concurrent requests. Each thread handles one request at a time, however more threads can be dynamically created to handle heavy loads. Creating the threads in advance reduces latency for handling incoming requests. Next the agent uses the *listen*() command to start listening on a predefined TCP port, such as 8080, for incoming connections.

When a new incoming client connection is received, Resumption Agent uses the connection's file descriptor to hand the request to a worker thread. The worker thread then uses asynchronous non-blocking UNIX IO, *read_nio*(),to read from the connection. It parses the client headers, then establishes a connection to that server and forwards the client's request headers to the server. When the server begins answering the request, the worker thread forwards the data back to the client. We note that for each transfer, Resumption Agent only has to process the header data; everything else is simply forwarded without processing overhead, minimizing any performance impact. Furthermore, in order to reduce CPU usage and bandwidth without increasing latency, the worker threads employ a large read size of 2KB, or the amount of data available in the system queue, from the server before forwarding it to the client.

The *socket*() implementation usually allows a socket timeout option to be specified, to report a disconnection if no data has been sent after the specified timeout. Unfortunately, while the iPhone appears to implement this option, it failed to function. Thus, we implemented our own timeout detection, with the same behaviour. If a timeout is detected, or the socket throws any error, the worker thread re-establishes a connection to the server and attempts to resume the transfer where it left off. The worker will retry up to a predefined number of times, by default 50, before giving up; this keeps the worker from running infinitely, and potential flooding the network, if the server or network becomes unavailable for an extended period of time.

We have measured the performance impact of Resumption Agent to be minimal in normal usage. In particular, Resumption Agent consumes less than 300 KB of memory, and it increases linearly with the number of concurrent transfers. Its mean CPU consumption is negligible when idle, and 3 – 4% when actively handling transfers. Most importantly, we have measured the additional latency introduced by the Resumption Agent to be statistically insignificant over 200 test runs.

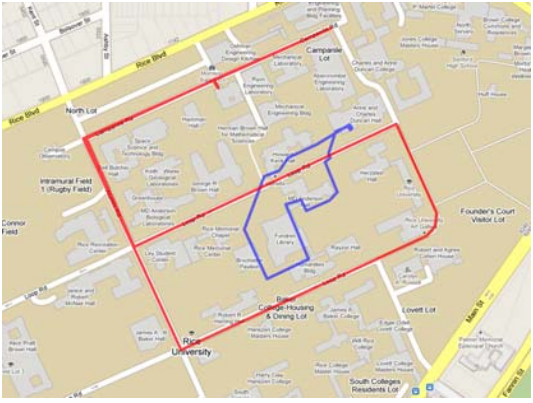

**Figure 16: Map of paths travelled in Rice University, for walking (1 km, Blue loop) and driving (3 km, Red loop) scenarios**

# 6. Example Application: AutoSwitch

In order to evaluate the combined effectiveness of the Wait-n-Migrate and Resumption Agent mechanisms, we have developed AutoSwitch, an automatic network interface switching policy. AutoSwitch attempts to offload data from cellular to Wi-Fi as much as possible, with minimum disruptions to the user. AutoSwitch solves a common complaint about Wi-Fi [40] – that it is unreliable or unusable at low signal levels, such as while the user is moving in and out of coverage areas. To achieve this goal, AutoSwitch intelligently switches between wireless networks using Wait-n-Migrate and Resumption Agent, before losing connectivity (e.g., due to mobility). We note that other solutions have been proposed to offload cellular traffic on Wi-Fi, e.g. [5], but they typically rely on a mobility gateway / proxy to handle network switches, with inherent latency and deployability drawbacks as discussed in Section 2.2.

## 6.1 AutoSwitch Design

AutoSwitch attempts to migrate TCP flows from Wi-Fi to cellular before Wi-Fi coverage is dropped or Wi-Fi becomes unreliable, and migrate back to Wi-Fi when a reliable Wi-Fi connection becomes available again. For simplicity, and without losing generality, we assume that cellular coverage is always available.

Often, in particular for the case of mobility, switching between networks occurs due to forced disconnections. For example, a phone may switch from 3G to a Wi-Fi network when Wi-Fi becomes available, but move out of Wi-Fi coverage shortly afterwards; thus the phone is forced to switch back to 3G. In such a case, it is too late to effectively use Wait-n-Migrate. However, previous work shows that it is indeed possible to accurately predict network conditions, and therefore initiate the network switch before losing coverage. For example, Breadcrumbs [41] and our previous work [42] predict network conditions for the near and far future, respectively. As our main focus is on flow migration and not on the switching policy, we use a simple yet effective predictor, signal strength [8, 43], to initiate a network switch before losing Wi-Fi coverage completely.

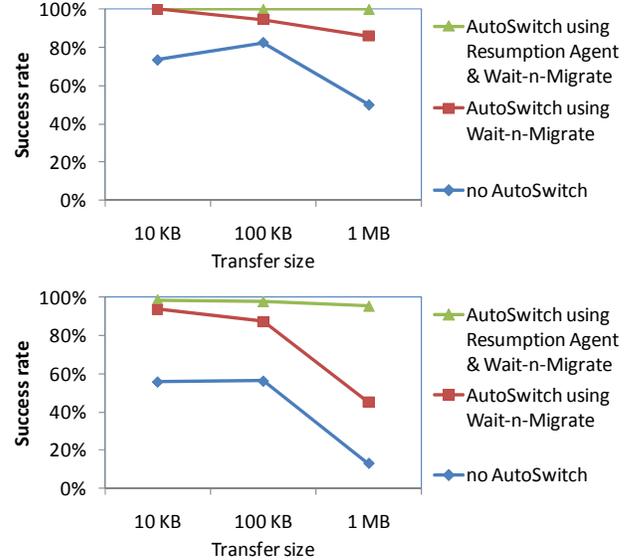

**Figure 17: AutoSwitch significantly increases the success rate of 10KB, 100KB, and 1MB transfers when walking (top) and driving (bottom) on Rice Campus**

In order to determine the policy for switching to and from Wi-Fi, we extended LiveLab for three iPhone 3GS users for three weeks to continuously test and record network disconnections, measured by the ping tool. These three users acted as a sampling tool to measure Wi-Fi reliability at different signal strengths, collecting over 1 million connectivity tests, shown in Figure 14. We define a Wi-Fi connection as disconnected if all ping tests over a period of 5 seconds are lost, regardless of the reported signal strength. We can see that Wi-Fi starts to become unreliable starting at approximately -82 dBm on iPhone 3GS.

Based on these results, we employ a simple hysteresis over both time and signal strength to reduce erroneous switching. AutoSwitch, using Wait-n-Migrate and Resumption Agent, switches to cellular when a Wi-Fi signal level of -75 dBm or less is maintained over 3 seconds, and switches back to Wi-Fi when Wi-Fi signal strength reaches -70 dBm.

## 6.2 Trace-based Evaluation

We have used the traces from LiveLab to evaluate the efficacy of AutoSwitch using Wait-n-Migrate, during routine interactive usage. As mentioned in Section 6.1, LiveLab provides us with continuous signal strength measurements, but not connectivity measurements. We utilize the Wi-Fi signal strength measurements and the probability of disconnection at different signal levels, presented in Figure 14, to calculate the expected number of disruption in a web application. We further assume that if Wi-Fi is not disconnected at a specific signal level in a particular usage session, it will not be disconnected at that signal level for the entire session.

We present the expected number of disconnections for AutoSwitch using Wait-n-Migrate, with wait-times of 10, 30, and 100 seconds, respectively, in Figure 15. We com-pare it with two cases, one where Wi-Fi is left on (no AutoSwitch), and another case where AutoSwitch switches between networks in a brute force manner, without utilizing Wait-n-Migrate.

Using 1120 hours of interactive usage traces with Wi-Fi enabled, for web usage, the users were expected to experience 213 disruptions without AutoSwitch. Employing AutoSwitch using Wait-n-Migrate and a constant wait-time of 10, 30, and 100 seconds, users were expected to experience 80%, 87%, and 91% fewer disconnections, respectively (Figure 15). In contrast, AutoSwitch with brute force switching, i.e., without Wait-n-Migrate, slightly increases disconnections to 246, due to false positives.

We must note that users indeed take note of the mobility and coverage limitations of Wi-Fi, as confirmed by our motivational user study from Section 2.1 and prior work [40]. Therefore, they may turn off Wi-Fi altogether in conditions they know it is prone to failing. Hence, we expect that the results in this section, obtained from the traces when Wi-Fi was enabled, underestimate the potential benefit from AutoSwitch using Wait-n-Migrate.

### 6.3 Field Evaluation

We further evaluate AutoSwitch using both Wait-n-Migrate and Resumption Agent on the iPhone platform. For performance evaluation we wrote a script to automatically download a predetermined file over HTTP, from a server that supports resuming, every five seconds. We tested AutoSwitch using transfer sizes of 10 KB, 100 KB, and 1MB, as well as Wait-n-Migrate alone. We then measured the number of transfers that were fully completed without errors over two predetermined paths in Rice University, shown in Figure 16; 1) while walking commonly used paths, and 2) while in a car travelling at approximately 30 km/h along campus roads. The walking path was approximately 1 km long, included indoor areas in two buildings, crossed distinct areas with good to excellent Wi-Fi connectivity (-70 dBm signal strength or more), and was covered approximately 95% of the time by a Wi-Fi signal. The driving path was approximately 3 km long and only had one area of good Wi-Fi signal strength, but still had about 80% Wi-Fi coverage. Each test run lasted approximately one hour, and included over 1000 transfer attempts.

The success rates of transfers, as observed by our script, are shown in Figure 17. As expected, due to Wi-Fi signal variations, there are a significant number of failed transfers without AutoSwitch. Using AutoSwitch in conjunction with Wait-n-Migrate significantly reduced the number of disruptions. Furthermore, since the server supported resuming, Resumption Agent, used in conjunction with Wait-n-Migrate, was able to further reduce disruptions, completely eliminating them while walking, and increased the success rate while driving to over 95–99% for different file sizes.

## 7. Discussion

Our work focuses on providing system mechanisms for migrating flows between networks. Various policies have been proposed to switch between or aggregate networks. AutoSwitch is one such policy, and unambiguously demonstrates the effectiveness of Wait-n-Migrate and Resumption Agent in supporting seamless flow migration. The system mechanisms we have presented here can also be utilized to enable the immediate deployment of many performance and efficiency-enhancing policies studied in the literature, without practical deployment issues:

*Multihoming / Load Balancing:* When used for load balancing and multihoming, Resumption Agent has the key advantage of knowing the length of a flow at its very early stages, through the HTTP response headers, as well as the properties and conditions of the available networks. This allows Resumption Agent to intelligently allocate each flow on the appropriate network interface.

*Striping:* Resumption Agent can support striping larger transfers, i.e. download different parts of the transfer simultaneously through different networks, as long as the content supports resuming. Resumption Agent can be extended to download separate chunks over each interface and then amalgamate these chunks before sending them to the client. For striping content that contains dynamic parts, as described in Section 4.2.1, it is necessary to ensure the dynamic portions are downloaded in single chunks.

*Mirroring:* For pages that do not support striping, or that are very small compared to the latency, Resumption Agent can be extended to simultaneously request the same page on multiple networks, and return whichever finishes first. While this method is not power-efficient, it can provide substantial reduction in user perceived latency, especially under highly varying network environments.

*Preemptive Network Switching:* When Resumption Agent is aware of an impending network switch, it can establish a connection over the new network and request the remaining portion of the flow, *before killing the existing flow*. This allows the Resumption Agent to further minimize the latency incurred when resuming a flow.

## 8. Conclusion

We presented a first-of-its-kind characterization of IP traffic on modern smartphones using traces collected in real-life usage of 27 iPhone 3GS users over a period of three months. We show that the traffic is almost exclusively TCP, and TCP flows are often short-lived and rarely concurrent for interactive applications.

Driven by these findings, we devised two novel and complementary system mechanisms to migrate TCP flows between networks without network or application support: Wait-n-Migrate and Resumption Agent. While Wait-n-Migrate significantly decreases, or even eliminates connectivity gaps when switching between networks,

Resumption Agent opportunistically resumes flows across connectivity disruptions and network switches. Combined, these two system mechanisms mitigate, and in many cases eliminate, the impact of widely varying network conditions on mobile applications, as we demonstrate using our implementation, AutoSwitch. The seamless flow migration without network support collectively enabled by Wait-n-Migrate and Resumption Agent allows for immediate deployment of performance and efficiency-enhancing policies, including multihoming and traffic offloading.